\documentclass[a4paper,11pt]{article}
\pdfoutput=1 % if your are submitting a pdflatex (i.e. if you have
\usepackage{jinstpub} % for details on the use of the package, please
\usepackage{multirow}
\usepackage[table]{xcolor}
\title{Cylindrical Films for Electronics in Low Background Physics Searches}

\author[a]{E. Brown,}
\author[a,1]{K. Odgers, \note{Corresponding author.}}
\author[a]{M. Giordano,}
\author[a, 2]{K. Lewis,\note{Presently: Department of Physics and Astronomy, College of Arts and Sciences, Howard University, Washington DC 20059, USA}}
\author[a]{T. Berger}
\author[b]{J. Freedberg}

\affiliation[a]{Department of Physics, Applied Physics and Astronomy, Rensselaer Polytechnic Institute, Troy NY, USA}
\affiliation[b]{Department of Physics, University of Minnesota, Minneapolis MN, USA}

\emailAdd{odgerk@rpi.edu}

\abstract{A technique for manufacturing thin-film resistors on cylindrical substrates is demonstrated. These devices are aimed for application in rare-event detectors that must minimize radioactive backgrounds from trace impurities in electronic components inside the detector. Cylindrical, conducting Ni films were created via Electron Beam Deposition, using a mechanism that rotates the substrate, to demonstrate proof of principle and measure the resistivity on axis and in azimuth.  These films are characterized by measurements using a facsimile of the Van Der Pauw method combined with electrostatic simulations. In the two cylindrical samples made we observe anisotropic electrical behavior with resistivities of 1392.5, 888.5 $n \Omega m$ around the azimuth and of 81.9, 72.8  $n \Omega m$ along the axis of the sample. We show that this anisotropy is not caused just by the electron beam evaporation by measuring a planar rectangle sample made in the same process but without spinning which has estimated resistivities of 66.5, and 71.9 $n \Omega m$ in both directions, and calculated resistivity using the standard Van der Pauw equation of $66.1\pm2.8$ $n \Omega m$. In spite of the anisotropy in the cylindrical samples, we show that these films can be used as resistors. }

\keywords{Dark Matter detectors (WIMPs, axions, etc.), Double-beta decay detectors, Time projection Chambers (TPC), Manufacturing}

\arxivnumber{} % only if you have one

\begin{document}
\maketitle
\flushbottom

\section{Introduction}
\label{sec:intro}

Thin film electronics with high radiopurity materials are an exciting prospect for rare event physics, and other experiments requiring a low radioactive background. In experiments measuring recoil energies from eV up to several MeV, trace radioimpurities in detector components contribute a substantial background for these experiments \cite{a} \cite{b}. To reach the low background needed for physics searches, each material used in the detectors must have a minimum of radio-contamination. The concentration of $^{238}$U and $^{232}$Th must be at or below the ppt level \footnote{1 ppt = $10^{-12}$mol/mol} \cite{a} \cite{h}\cite{pmt}\cite{i}.

One class of rare event detectors uses a time projection chamber (TPC) filled with a cryogenic liquid \cite{g}. There are two types of TPCs, either single or dual phase. These correspond to whether the detecting material is entirely liquid, or if it has liquid and gas segments. The TPC is based off a particle interaction or decay in the liquid phase yielding two signals. This radiation induces a prompt scintillation signal that is measured by photosensors, and an ionization signal. The ionized electrons are drift through the cryogenic liquid to an anode via a strong electric field where a charge signal is read out. The time delay of the two signals and localization of the readout of the ionization signal allow 3D position reconstruction. 

To generate the uniform drift field in a liquid noble TPC, field shaping electrodes line the outer edge of the detector with a cascading potential. The cascade requires a resistor chain, and the shaping electrodes require electrically insulating structural support. Even small resistors contribute sizably to radioactive backgrounds, as do the macroscopic quantities of plastics typically used to support the electrodes. Thus, future detectors are focusing efforts to alleviate these sources of background.

One strategy for minimizing electronics near the fiducial region is to use small resistive spacers to link the field shaping electrodes. These can be realized by depositing a resistive film on the surface of insulating rods. The high purity and small quantity of material deposited for a film on the order of hundreds of nanometers allows the construction of highly radiopure support rods. To this end, we developed a system for depositing films onto cylinders, which can be used to build resistive electrode supports. We present here a proof of concept for the production and characterization of these cylindrical films.

Four point probe measurements are a common practice for measuring the resistivity of planar sample, and techniques commonly use either a linear series of probes, or probes attached to four corners of a rectangular sample \cite{e}. Probe locations analogous to this method were used to characterize the cylindrical samples studied and results were compared to simulations of the devices.

\section{Film Production and Rotation Device}

\begin{figure}[htbp]
\centering % \begin{center}/\end{center} takes some additional vertical space
\includegraphics[width=.4\textwidth]{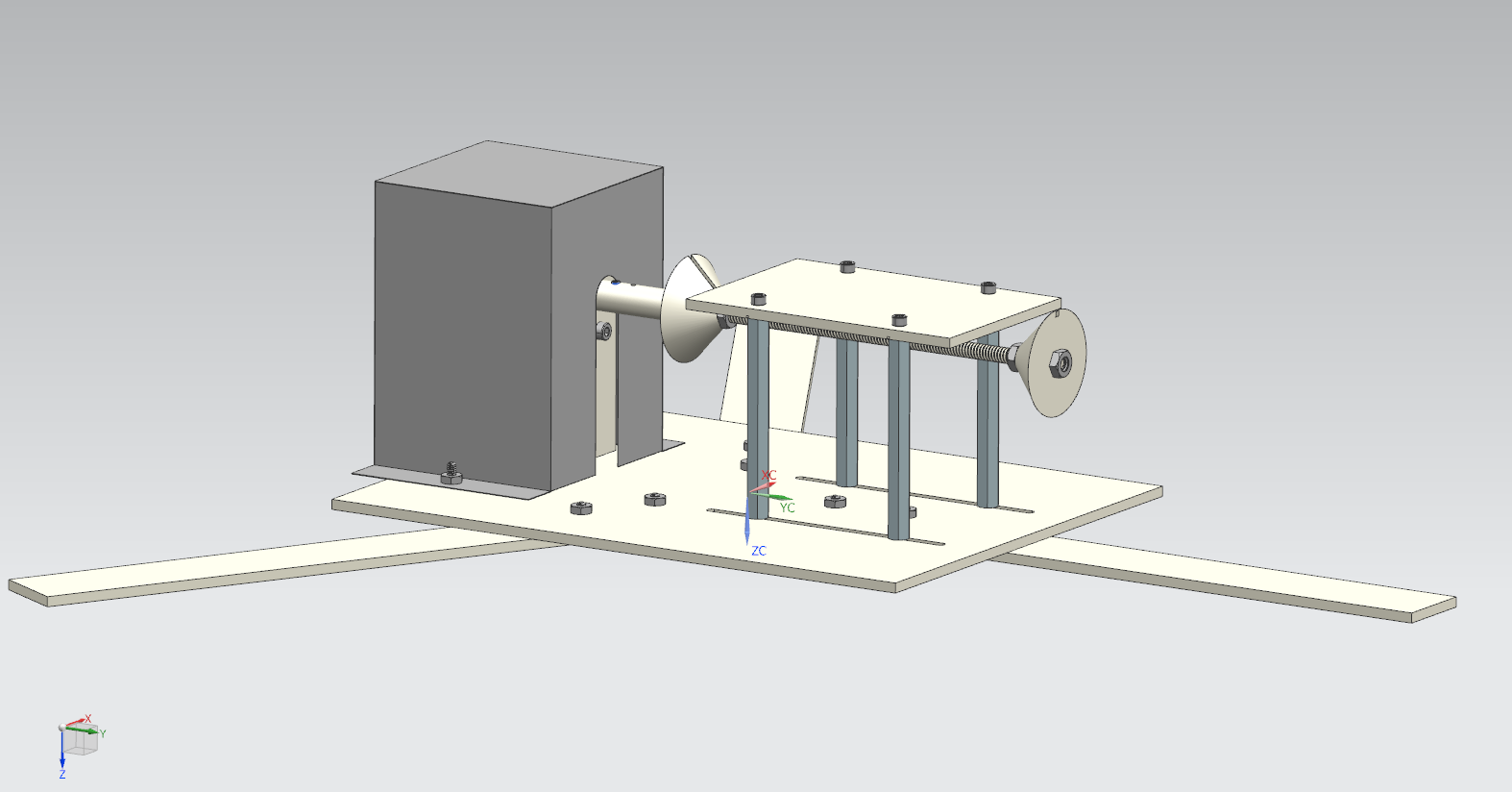}
\qquad
\includegraphics[width=.3\textwidth, angle = 0]{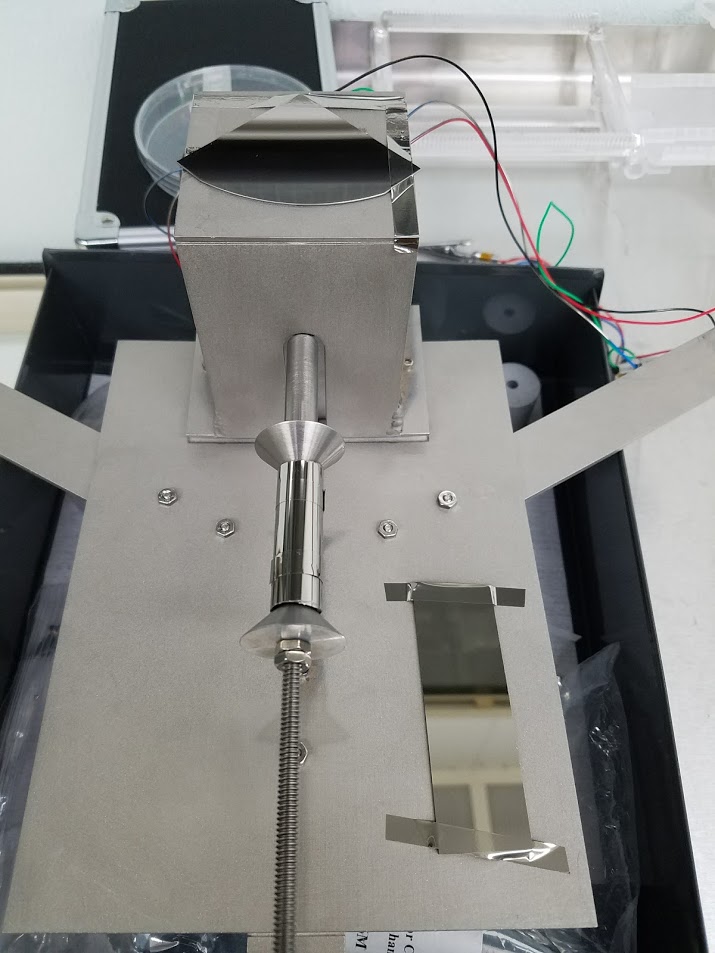}
% "\includegraphics" from the "graphicx" permits to crop (trim+clip)
% and rotate (angle) and image (and much more)
\includegraphics[width=.2\textwidth, angle = 90]{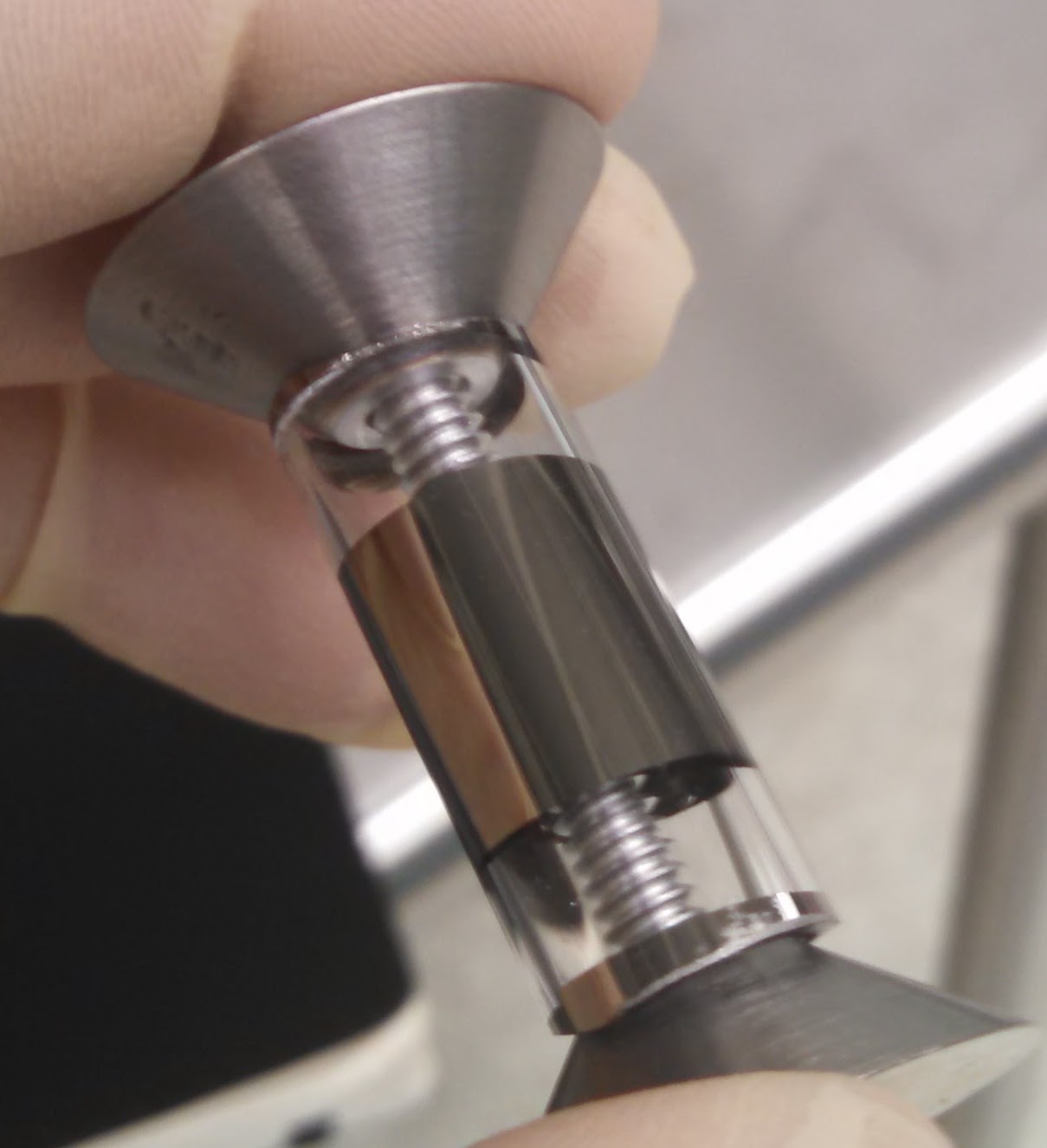}
\caption{\label{CFM}Left) The schematic of the motorized substrate holder with the arms, motor body and  substrate mount and mask. Center) Fully constructed device with several different substrates attached prior to implementing the mask.  Right) Cylindrical sample after having the kapton tape removed showing the clean edges and that the film being deposited and measured is far from where scrapes and scores would occur.}
\end{figure}

Films are prepared in a Temescal Electron Beam Evaporator. To produce cylindrical films, a mechanism was built to hold and turn a cylindrical substrate in the vacuum environment as seen in figure \ref{CFM}. The mechanism was built to hold a hollow tube ranging from 2 to 10 centimeters in length in the middle of the vacuum chamber, parallel to the bottom of the chamber. The extra space on the mechanism allows mounting of additional planar samples that can be used as a calibration for each deposition by measuring the thickness and resistivity of planar samples produced simultaneously with the cylindrical samples.

When rotating the substrates during deposition, some corrections must be made to interpret the thickness measurement of the in-situ quartz crystal resonance instrument. The deposition covered a substrate area of $2 \pi r l$ however the cross sectional area of the substrate during the deposition was $2 r l$. A correction of $\frac{1}{\pi}$ for the film thickness on the cylinder compared to that measured by the crystal and planar samples must be applied. Additionally, a mask was installed to restrict the deposition width to 3 mm, which further reduced the thickness of the cylindrical sample versus planar, the exact factor of which depends on the outer diameter of the tube.

The cylindrical quartz substrate was cut from a long tube with  an outer diameter of 12.7 mm, with a substrate length on the order of several centimeters. Cutting the tube left scoring near the top and bottom of the sample, and uneven edges, which were addressed after cleaning. The cleaning and preparation for the quartz cylinders, quartz slides, and silicon wafer was a piranha etch of equal parts sulfuric acid and hydrogen peroxide. Afterwards, the sample was rinsed with distilled water and dried with nitrogen. Kapton tape was used on the cylindrical substrate to mitigate the jagged and scored ends from the cutting process and ensure clean, parallel edges at the two ends of the substrate, as can be seen in figure \ref{CFM}.

A nickel source was used for the deposition onto the bare quartz. For any nickel film thicker than 40 nanometers the conductivity of the samples is too high for use as a resistor, but for developing the process, choosing a metal which would not have any Schottky barrier or diodic behavior was important. Our conclusions, however, can be applied to more resistive materials. The entire assembly was put into the deposition chamber and pumped down to $5\times 10^{-7}$ Torr before deposition was started. After heating the nickel source to the predetermined electron beam power for deposition, the cylindrical substrate holder was set to 60 RPM and the shutter between the nickel and substrate was opened. The deposition was maintained between 1-3 nm/s for the entire duration of the deposition as measured by the in-situ crystal resonance instrument. The crystal monitor measures the total thickness in-situ as well as deposition rate in angstroms per second for a planar sample. Due to the planar to cylindrical correction factor discussed above, the target thickness as read by the monitor was increased.

The first nickel sample examined was manufactured without a mask. Therefore, half of the surface of the cylinder was exposed to the nickel source at any given time during the deposition. A second sample was made with a mask that had a 3mm cut out parallel to the axis of the substrate, so the film was deposited onto the "flattest" portion of the substrate with the total deposition thickness being varied as well.

\section{Measurement Method}
\subsection{Flat Substrate Calibration}
Measurements of the planar samples were made with a Keithley 2400 Source Meter as a four point probe device. The probes were placed at four corners of a small, rectangular cleaved portion of the nickel-on-wafer sample. Silver paint was used to connect copper wires to the designated contact areas which measured no more than 10 square millimeters. The current was applied between all possible combinations of two adjacent contacts (in other words, not including diagonals). The other two contacts not supplied with a current had the potential difference between them measured. These are standard measurements for the Van Der Pauw Method which uses equation \ref{vanderpauw} to calculate resistivity from the currents and voltages \cite{vdp}

\begin{equation}\label{vanderpauw}
e^{\frac{-\pi c V_{H}}{I_0 \rho}}+e^{\frac{-\pi c V_{V}}{I_0 \rho}} =1
\end{equation}
In equation \ref{vanderpauw}, we use the potential differences in the horizontal and vertical edge directions ($ V_{H} , V_{V}$), the film thickness ($c$), and the supplied current ($I_0$) to fit the resistivity ($\rho$). The thickness is known from the crystal resonance measurement made during deposition of the film. The last two variables, current and voltage, were set and measured by the Keithley Source Meter.

Equation \ref{vanderpauw}, can be modified and used for most samples which are topologically equivalent (homeomorphic) to a rectangle. For topographically different samples, modifications to the equation must be made as is discussed in other papers \cite{cylvan}. The cylindrical samples fabricated for this study can be described as topographically different from the closed surface required to use the Van Der Pauw equation. When mapped to a flat surface, the cylinder would be a surface with a hole in it. For these purposes, equation \ref{vanderpauw} is not used in determining the resistivity of the samples and instead a different technique is introduced in section \ref{Sim}.

\subsection{Cylindrical Four Point Measurement}

\begin{figure}[htbp]
\centering % \begin{center}/\end{center} takes some additional vertical space
\includegraphics[width=.8\textwidth,clip]{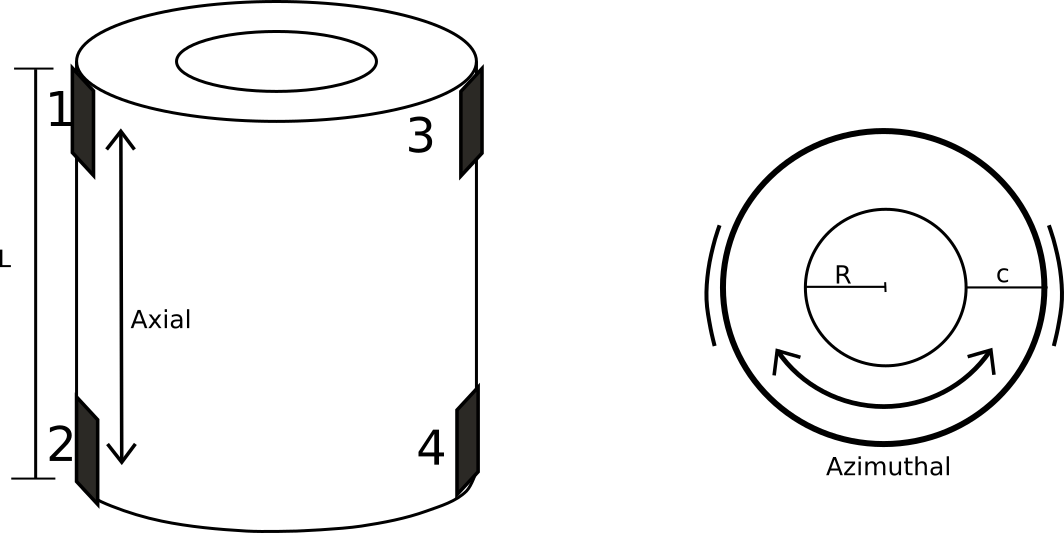}

\caption{\label{Geometry} Geometry of Film on Glass and Probe/Source Locations}
\end{figure}

Similar to the Van Der Pauw method used for the planar sample, four probes were attached to each cylindrical sample near what could be considered corners. Two were placed at the top and bottom of the sample in a line parallel with the axis of the cylinder, and on the other side (rotated $\pi$ radians) the other two were placed in a similar position. This can be seen in figure \ref{Geometry}. %How then do we extract the resistivity? We have to describe that here
For both cylindrical samples, measurements similar to those made of the planar sample were made. After room temperature measurements, one of the sample cylinders was placed in a cryostat, and measurements were repeated as temperatures were cycled from 100 K to 300 K as seen in figure \ref{Temp}.

\subsection{\label{Sim}Electronic Simulations of Cylindrical Films}

In conjunction with the construction and measurement of the films, a finite element model of the film was made in COMSOL 5.1  with the Electric Current Shell module (ECS) \cite{d}. Simulations were made for both the planar and cylindrical samples. For the planar model, two current sources (one positive and one negative) were placed on the modeled film along with two voltage probes as well. The current sources were set to 0.5 and -0.5 mA respectively. The resistivity was left as a free parameter to be varied for each solution. The difference between voltages probes was recorded for many iterations of the simulation with resistivities spanning an order of magnitude above and below what is expected for nickel. These values were used to find a function for voltage difference given some resistivity using linear regression. These functions were found for each orientation of probes (across the width and length) and then inverted to obtain a formula for resistivity given a measured voltage. Applying the inverted formula to the actual voltage measurements is how the resistivities reported were determined.

The ECS module was used for the cylindrical film models as well, with a similar process employed to construct a resistivity formula. A solid which accurately reflected the dimensions of the cylindrical films was generated in the software, with four regions defined at the points where the contacts were painted on physical samples. The voltage at the two probe simulations was then simulated at various values of resistivity. The same fitting and inverting was applied to get the resistivity formulas for each probe orientation for each cylindrical sample. Each of these formulas was applied to the corresponding measured potential differences to determine the reported resistivities of the samples.

\section{Results}
In table \ref{Measurements}, we report measurements made for the two cylindrical samples as well as the planar sample.  The reported polar and axial voltages and resistivities would be those where currents were passed from $1 \rightarrow 3$ and $1 \rightarrow 2$ as labelled in figure \ref{Geometry}. The voltages reported are averages of four separate but complimentary measurements in the same orientation. For instance, the reported axial $\Delta V$ comes from averaging $V_{1,2},V_{2,1},V_{3,4}$ and $V_{4,3}$, corresponding to a measurement, its reversed polarity, its reciprocal, and its reciprocal in reversed polarity. Averaging these voltage measurements is standard practice to reduce the effect of unwanted voltage biases. The resistivities presented are those estimated using the comparison to COMSOL as detailed in section \ref{Sim} for the voltages reported.

\begin{table}[htbp]
\centering
\caption{\label{Measurements} Voltages for orthogonal probe orientations with 1 mA source. For the cylindrical samples, Directions 1 (D1) and 2 (D2) correspond to polar and axial measurements of the sample respectively. For the planar, the directions correspond to measurements along the width and length of the rectangular sample. For both types of sample, C is the thickness of the sample.}
\smallskip
\begin{tabular}{|c|c|c|c|c|c|c|}
\hline
Film &R x L [mm]& C [nm] & D1 $\Delta V$[V] & D2 $\Delta V$[V] & D1 $\rho$ [$n\Omega m$] & D2 $\rho$ [$n\Omega m$]\\
\hline
Cyl 1  & 6.4 x 18 & 80 & 1.301 & 0.044 &1392.5&81.9\\
Cyl 2  &6.4 x 17& 31 & 2.517 & 0.064&888.5&72.8\\
\hline
\multicolumn{7}{|c|}{}\\
\hline
 \cellcolor{black!25}  &  WxL [mm] & C [nm] &\multicolumn{4}{|c|}{\cellcolor{black!25}}\\

\hline
Planar &6x15& 120 & .449 & .014 & 66.5 & 71.9\\
%%\hline
%%Planar &\multicolumn{4}{|c|}{Van Der Pauw Calculated} & 66.1$\pm$2.8 & \\
\hline
\end{tabular}
\end{table}

The accepted resistivity of nickel depends on the purity of the nickel as well as what contaminants are in the sample \cite{resistrange}. While some nickel samples can have a resistivity as high as 167 $n\Omega m$ , we compare our measurements to a value of 61.6 $ n\Omega m$ as given by the National Physics Laboratory in the UK \cite{f}. The ratio of resistivities for directions D1 to D2 is nearly unity for the planar sample (1.09), but is more than an order of magnitude larger for both of the cylindrical samples(17.0 and 12.2). We interpret this as an electrical anisotropy for the cylindrical samples as described in section 4.1.When the axial direction of the cylinder was probed the biggest relative difference was 18\% between measured and accepted resistivity. The polar resistivities disagreed strongly with the accepted value with a relative error of 1900\%.

For the planar sample, equation \ref{vanderpauw} was used to calculate the bulk resistivity as a check to show that the method of fitting to COMSOL simulations agreed with accepted practices. When this standard analysis was performed, the resistivity of the sample found was $66.1 \pm 2.8$ $n\Omega m$. Equation \ref{vanderpauw}, returns a single resistivity for the sample, unlike our comparison to COMSOL, but when we compare the calculated value to either of our fitted values, or to the accepted resistivity of nickel we observe the values are all roughly in agreement. The uncertainty on the resistivity calculated with equation \ref{vanderpauw} was calculated by adding the uncertainty on voltage and thicknesses in quadrature. Because of the transcendental nature of equation \ref{vanderpauw}, numerical derivatives were taking with a 5-point stencil for calculating the error. 

Because of the nature of our fitting to simulation, uncertainties and errors were left off for the fitted values. Possible sources of error could be a non-uniform deposition, an incorrect correction for thickness on the cylinder, contact resistance with the silver used for bonding the probes to the film, or simulation errors. The low relative error for both planar measurements shows that this method was successful within a small factor. This holds true for the estimated resistivity for the axial probe orientation, but not for the polar orientation. This discrepancy is discussed later.

\begin{figure}[htbp]
\centering % \begin{center}/\end{center} takes some additioresistancenal vertical space
\includegraphics[width=.8\textwidth,clip]{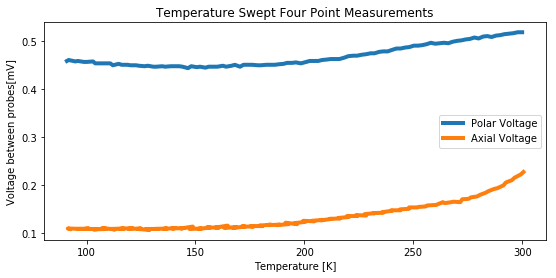}

\caption{\label{Temp} Low Temperature Performance of a Thin Cylindrical Film }
\end{figure}

The last measurement completed was the same four point measurement but performed as the temperature warmed from 100 K to 300 K. In figure \ref{Temp} it is shown that the voltage difference of the substrate and thus resistance increases as the temperature rises. This is consistent with what is expected for a metal, and shows that electrical properties of these cylindrical films can return to normal after heating and that the films survive cold temperatures and thermal contractions and expansions of the quartz. This is a critically important feature for applications in cryogenic liquid detectors.

\subsection{Electrical Anisotropy}
Table \ref{Measurements} shows that the resistivity found by axial and polar probe measurements differed by more than an order of magnitude. The simulations used to estimate the resistivity to voltage relationship used isotropic resistivities, so the numbers in table \ref{Measurements} do not represent the physical vector components of the resistivity. However, the resistivity definitively shows that there is an anisotropy to the film, and that resistivity is greater going around the circumference of the cylinder versus along the axial direction. The mechanism of this anisotropy has not been explored but several possible explanations have been formed. The most likely cause is that the middle of the substrate during the deposition is perpendicular to the direction, but as you deviate from the middle the angle between substrate and incoming particles becomes oblique. Glancing Angle Deposition has been used to form meso and nano structures, and may play a part in this \cite{c}. It is also possible that grain growth is easier along the axis of the cylinder, giving the electrical bias in that direction.

While the source of the anisotropy observed in the resistivity is of interest, for the purposes of constructing resistors from thin films it should have a negligible impact if the currents are run along the axial direction of the cylinder. This is due to the fact that a bulk resistivity is sufficient for operating a thin film resistor in applications for liquid noble TPCs. We thus leave precise investigations of the source, impact and potential use of these effects to further studies.

% "source, impact and potential use of these effects"?
%sure

% \begin{figure}[htbp]
% \centering % \begin{center}/\end{center} takes some additional vertical space
% \includegraphics[width=.4\textwidth]{isotop}
% \qquad
% \includegraphics[width=.4\textwidth]{antop}
% % "\includegraphics" from the "graphicx" permits to crop (trim+clip)
% % and rotate (angle) and image (and much more)
% \caption{\label{Simulation} Comsol Simulation with Effect of Isotropic Resistivity versus Anisotropic with Half the Axial Resistivity}
% \end{figure}

\section{Conclusion}

 We have successfully demonstrated the ability to produce thin, resistive films on cylindrical surfaces for use in cryogenic applications. Our proof of concept using nickel showed the reproducibility of our fabrication method, and through comparison to COMSOL simulations we can characterize the bulk resistivity of our samples using a four point probe method of measurement. For the planar sample, calculation from equation \ref{vanderpauw} yielded a resisitvity of $66.1 \pm 2.8$ $n\Omega m$, while the COMSOL simulations yielded $66.5$ and $71.9$ $n\Omega m$ for the two directions. This is in fair agreement with the literature value of $61.6$ $n \Omega m$ \cite{f}. In characterizing the cylindrical samples through comparison to simulations with an isotropic resistivity, the resistivity for the axial and polar probe orientations differed by more than an order of magnitude. For the two samples, the respective resitivities were 81.9 and 1392.5 $n \Omega m$ and 72.8 and 888.5 $n \Omega m$. This difference when fitting the resistivities leads to the conclusion that there is an anistotropic resistivity for these cylindrical samples, and that using a mask can affect this anisotropy. This hints to the cause of the anisotropy being the glancing angles during deposition. Stable operation in the temperature range of 77 to 300 K show promise for creating thin film resistors of this type using other materials for use in cryogenic applications.

\section{Acknowledgments}

We would like to thank Poomirat Nawarat for his time and help in finalizing this study.


\end{document}